\documentclass[aps, prd, floatfix, nofootinbib, superscriptaddress, twocolumn]{revtex4-1}

\usepackage{latexsym}
\usepackage{amsmath}
\usepackage{amssymb}
\usepackage{amsfonts}

\usepackage[mathscr,scaled=1.15]{urwchancal}
\DeclareFontFamily{OT1}{pzc}{}
\DeclareFontShape{OT1}{pzc}{m}{it}%
{<-> s * [1.15] pzcmi7t}{}
\DeclareMathAlphabet{\mathpzc}{OT1}{pzc}{m}{it}

\usepackage{color}

\usepackage{supertabular}
\usepackage{placeins}
\usepackage{epsfig}
\usepackage{graphicx}

\definecolor{purple}{rgb}{0.5,0,0.5}
\definecolor{blue}{rgb}{0.0,0,0.9}
\definecolor{prdblue}{rgb}{0.133,0.118,0.498}
\usepackage[colorlinks=true, pdfstartview=FitV, linkcolor=prdblue, citecolor= prdblue, urlcolor=prdblue]{hyperref}

\begin{document}


\title{$\,$\\[-7ex]\hspace*{\fill}{\normalsize{\sf\emph{Preprint no}. NJU-INP 012/19}}\\[1ex]
Drawing insights from pion parton distributions}

\author{Minghui Ding}
\affiliation{School of Physics, Nankai University, Tianjin 300071, China}
\affiliation{European Centre for Theoretical Studies in Nuclear Physics
and Related Areas (ECT$^\ast$) and Fondazione Bruno Kessler\\ Villa Tambosi, Strada delle Tabarelle 286, I-38123 Villazzano (TN) Italy}

\author{Kh\'epani Raya}
\affiliation{School of Physics, Nankai University, Tianjin 300071, China}

\author{Daniele Binosi}
\affiliation{European Centre for Theoretical Studies in Nuclear Physics
and Related Areas (ECT$^\ast$) and Fondazione Bruno Kessler\\ Villa Tambosi, Strada delle Tabarelle 286, I-38123 Villazzano (TN) Italy}

\author{Lei Chang}
\email[]{leichang@nankai.edu.cn}
\affiliation{School of Physics, Nankai University, Tianjin 300071, China}

\author{C.~D.~Roberts}
\email[]{cdroberts@nju.edu.cn}
\affiliation{School of Physics, Nanjing University, Nanjing, Jiangsu 210093, China}
\affiliation{Institute for Nonperturbative Physics, Nanjing University, Nanjing, Jiangsu 210093, China}

\author{S. M. Schmidt}
\affiliation{
Institute for Advanced Simulation, Forschungszentrum J\"ulich and JARA, D-52425 J\"ulich, Germany}


\date{25 November 2019}   

\begin{abstract}
A symmetry-preserving continuum approach to the two valence-body bound-state problem is used to calculate the valence, glue and sea distributions within the pion; unifying them with, \emph{inter alia},  electromagnetic pion elastic and transition form factors.  %
The analysis reveals the following momentum fractions at the scale $\zeta_2:=2\,$GeV:
$\langle x_{\rm valence} \rangle = 0.48(3)$,
$\langle x_{\rm glue} \rangle  = 0.41(2)$,
$\langle x_{\rm sea} \rangle  = 0.11(2)$;
and despite hardening induced by the emergent phenomenon of dynamical chiral symmetry breaking, the valence-quark distribution function, ${\mathpzc q}^\pi(x)$, exhibits the $x\simeq 1$ behaviour predicted by quantum chromodynamics (QCD).  After evolution to $\zeta=5.2\,$GeV, the prediction for ${\mathpzc q}^\pi(x)$ matches that obtained using lattice-regularised QCD.  This confluence should both stimulate improved analyses of existing data and aid in planning efforts to obtain new data on the pion distribution functions.
\end{abstract}

\maketitle


\noindent\emph{1.$\;$Introduction} ---
%
Regarding their valence quark content, pions are Nature's simplest hadrons: $\pi^+  \sim u\bar d$, $\pi^- \sim d \bar u$, $\pi^0 \sim u\bar u - d\bar d$;
but this appearance is misleading.  Despite being hadrons, their physical masses are similar to that of the $\mu$-lepton; and the pion masses vanish in the absence of a Higgs coupling for $u$- and $d$-quarks.  Pions are Nambu-Goldstone (NG) modes, generated by dynamical chiral symmetry breaking (DCSB) in the Standard Model.  This dichotomous character -- simultaneous existence as both NG-bosons and bound-states -- entails that the challenges of charting and explaining pion structure are of central importance in modern physics \cite{Geesaman:2015fha}.  These problems are made more difficult by the crucial role of symmetries and their breaking patterns in determining pion properties, which must be properly incorporated and veraciously expressed in any theoretical treatment.

Given their simple valence-quark content, a basic quantity in any discussion of pion structure is the associated distribution function, ${\mathpzc q}^\pi(x;\zeta)$.
This density charts the probability that a valence ${\mathpzc q}$-quark in the pion carries a light-front fraction $x$ of the system's total momentum when the observation is made at resolving scale $\zeta$; and one of the earliest predictions of the parton model, augmented by features of perturbative quantum chromodynamics (pQCD), is \cite{Ezawa:1974wm, Farrar:1975yb, Berger:1979du}:
\begin{equation}
\label{PDFQCD}
{\mathpzc q}^{\pi}(x;\zeta =\zeta_H) \sim (1-x)^{2}\,,
\end{equation}
where the energy scale $\zeta_H$ marks the boundary between strong and perturbative dynamics.  Moreover, the exponent evolves as $\zeta$ increases beyond $\zeta_H$, becoming $2+\gamma$, where $\gamma\gtrsim 0$ is an anomalous dimension that increases logarithmically with $\zeta$.  (In the limit of exact ${\mathpzc G}$-parity symmetry, which is a good approximation in the Standard Model, $u^{\pi^+}(x) = \bar d^{\pi^+}(x)$, etc. Hence it is only necessary to discuss one unique distribution.)

${\mathpzc q}^\pi(x)$ is measurable in $\pi$-nucleon Drell-Yan experiments \cite{Badier:1980jq, Badier:1983mj, Betev:1985pg, Falciano:1986wk, Guanziroli:1987rp, Conway:1989fs, Heinrich:1991zm}. However, conclusions drawn from analyses of these experiments have proved controversial \cite{Holt:2010vj}.
For instance, using a leading-order (LO) pQCD analysis of their data, Ref.\,\cite{Conway:1989fs} (the E615 experiment) reported 
\begin{align}
{\mathpzc q}_{\rm E615}^{\pi}(x; \zeta_5=5.2\,{\rm GeV}) &\sim (1-x)^{1}\,,
\end{align}
in conflict with Eq.\,\eqref{PDFQCD}.  Subsequent calculations \cite{Hecht:2000xa} confirmed Eq.\,\eqref{PDFQCD}, prompting reconsideration of the E615 analysis, with the result that, at next-to-leading order (NLO) and including soft-gluon resummation \cite{Wijesooriya:2005ir, Aicher:2010cb}, the E615 data become consistent with Eq.\,\eqref{PDFQCD}.

Notwithstanding these advances, uncertainty over Eq.\,\eqref{PDFQCD} will remain until other analyses of the E615 data incorporate threshold resummation effects and, crucially, new data are obtained. Prospects for the latter are good because relevant tagged deep-inelastic scattering experiments are approved at the Thomas Jefferson National Accelerator Facility \cite{Keppel:2015, Keppel:2015B, McKenney:2015xis} and the goal has high priority at other existing and anticipated facilities \cite{Petrov:2011pg, Peng:2016ebs, Peng:2017ddf, Horn:2018fqr, Denisov:2018unj, Aguilar:2019teb}.

Meanwhile, progress in theory continues.
Novel algorithms within lattice-regularised QCD (lQCD) \cite{Liu:1993cv, Ji:2013dva, Radyushkin:2016hsy, Radyushkin:2017cyf, Chambers:2017dov} are beginning to yield results for the pointwise behaviour of the pion's valence-quark distribution \cite{Xu:2018eii, Chen:2018fwa, Karthik:2018wmj, Sufian:2019bol}, promising information beyond the lowest few moments \cite{Best:1997qp, Detmold:2003tm, Brommel:2006zz, Oehm:2018jvm}.
Extensions of the continuum analysis in Ref.\,\cite{Hecht:2000xa} are also yielding new insights.  For example:
a class of corrections to the handbag-diagram representation of the virtual-photon--pion forward Compton scattering amplitude has been identified and shown to restore basic symmetries in calculations of ${\mathpzc q}^\pi(x;\zeta)$ \cite{Chang:2014lva}.
The corrected expression has been used to compute all valence-quark distribution functions in the pion and kaon \cite{Chen:2016sno}, with the results indicating that the pion's gluon content is significantly greater than that of the kaon owing to the mechanism responsible for the emergence of mass in the Standard Model.
%

%

%
Capitalising on such recent developments, herein we describe predictions for the pion's parton distributions using a continuum approach that has been used successfully to unify the treatment of the charged-pion-elastic and neutral-pion-transition form factors \cite{Chang:2013nia, Raya:2015gva, Raya:2016yuj, Gao:2017mmp, Ding:2018xwy}.  The framework has also been used to correlate continuum and lattice predictions for the electromagnetic form factors of charged pion-like mesons,  enabling an extrapolation of lQCD results to the physical pion mass \cite{Chen:2018rwz}.

\smallskip

\noindent\emph{2.$\;$Valence-quark distribution function} ---
%
%
Incorporating the impact of global symmetries and DCSB, we compute the pion's valence-quark distribution function using the following expression \cite{Chang:2014lva}:
\begin{subequations}
\label{qFULL}
\begin{align}
{\mathpzc q}^\pi&(x;\zeta_H)  = N_c {\rm tr}\! \int_{dk}\! \delta_n^{x}(k_\eta) \,
{\mathpzc I}^\pi(k;P;\zeta_H)\,,\\
{\mathpzc I}^\pi & = n\cdot\partial_{k_\eta} \left[ \bar\Gamma_\pi(k_\eta,-P) S(k_\eta) \right]
\Gamma_\pi(k_{\bar\eta},P)\, S(k_{\bar\eta})\,,
\end{align}
\end{subequations}
where
$N_c=3$;
the trace is over spinor indices;
$\int_{dk} := \int \frac{d^4 k}{(2\pi)^4}$ is a translationally invariant regularisation of the integral;
$\delta_n^{x}(k_\eta):= \delta(n\cdot k_\eta - x n\cdot P)$;
$n$ is a light-like four-vector, $n^2=0$, $n\cdot P = -m_\pi$;
and $k_\eta = k + \eta P$, $k_{\bar\eta} = k - (1-\eta) P$, $\eta\in [0,1]$.
${\mathpzc q}^\pi(x;\zeta_H)$ in Eq.\,\eqref{qFULL} is independent of $\eta$; satisfies baryon number conservation; and is symmetric:
\begin{equation}
\label{xoneminusx}
{\mathpzc q}^\pi(x;\zeta_H) = {\mathpzc q}^\pi(1-x;\zeta_H)\,.
\end{equation}

To calculate ${\mathpzc q}^\pi$ from Eq.\,\eqref{qFULL} one needs the dressed light-quark propagator, $S$, and pion Bethe-Salpeter amplitude, $\Gamma_\pi$.  We follow Ref.\,\cite{Chen:2018rwz} and use realistic results calculated with renormalisation-group-invariant current-quark mass
$\hat m_u = \hat m_d =  6.7\,{\rm MeV}$\,,
which corresponds to a one-loop evolved mass of $m^{\zeta_2=2\,{\rm GeV}} = 4.6\,$MeV.  Consequently, the result for ${\mathpzc q}^\pi$ is completely determined once a kernel is specified for the Bethe-Salpeter equation.  We use the interaction explained in Refs.\,\cite{Qin:2011dd, Qin:2011xq}, whose connection with QCD is described in Ref.\,\cite{Binosi:2014aea}.  In solving all integral equations relevant to the bound-state problem, we employ a mass-independent momentum-subtraction renormalisation scheme and renormalise at $\zeta=\zeta_H$.

The value of $\zeta_H$ must now be determined.  To that end, recall that QCD possesses a process-independent effective charge \cite{Binosi:2016nme, Rodriguez-Quintero:2018wma}: $\alpha_{\rm PI}(k^2)$.  This running-coupling saturates in the infrared: $\alpha_{\rm PI}(0)/\pi \approx 1$, owing to the dynamical generation of a gluon mass-scale \cite{Boucaud:2011ug, Aguilar:2015bud}.  These features and a smooth connection with pQCD are expressed via
\begin{align}
\label{alphaPI}
\alpha_{\rm PI}(k^2) = \frac{\pi \gamma_m  }{\ln[(m_\alpha^2+k^2)/\Lambda_{\rm QCD}^2]}\,,
\end{align}
$m_\alpha = 0.30\,$GeV$\,\gtrsim \Lambda_{\rm QCD}$, QCD's renormalisation-group-invariant mass-scale: $\Lambda_{\rm QCD}\approx 0.23\,$GeV with four active quark flavours.  Evidently, $m_\alpha$ is an essentially nonperturbative scale whose existence ensures that modes with $k^2 \lesssim m_\alpha^2$ are screened from interactions.  It therefore serves to define the natural boundary between soft and hard physics; hence, we identify
$\zeta_H=m_\alpha$\,.

Using numerical solutions for $S$ and $\Gamma_\pi$, one can calculate the Mellin moments:
\begin{subequations}
\label{MellinMoments}
\begin{align}
& \langle   x^m  \rangle_{\zeta_H}^\pi  = \int_0^1dx\, x^m {\mathpzc q}^\pi(x;\zeta_H) \label{MellinA}\\
& = \frac{N_c}{n\cdot P} {\rm tr}\! \int_{dk}\! \left[\frac{n\cdot k_\eta}{n\cdot P}\right]^m {\mathpzc I}^\pi(k;P;\zeta_H)\,;
%
\end{align}
\end{subequations}
and if enough of these moments are computed, then they can be used to reconstruct the distribution.  Using Eq.\,\eqref{xoneminusx}, one finds that the value of any given odd moment, $\langle x^{m_{\rm o}}\rangle_{\zeta_H}^{\pi}$, $m_{\rm o}=2 \bar m +1$, $\bar m \in \mathbb Z$, is known once all lower even moments are computed.  Consequent identities can be used to validate any numerical method for computing the moments defined by Eq.\,\eqref{MellinMoments}.

Every moment defined by Eq.\,\eqref{MellinMoments} is finite.  However, direct calculation of the $m\geq 3$ moments using numerically determined inputs for $S$, $\Gamma_\pi$ is difficult in practice owing to an amplification of oscillations produced by the $[n\cdot k_\eta]^m$ factor.  In any perfect procedure, the oscillations cancel; but that is difficult to achieve numerically.  On $m\geq 3$, we therefore introduce a convergence-factor,
${\mathpzc C}_m(k^2 r^2)= 1/[1+k^2 r^2]^{m/2}$:
the moment is computed as a function of $r^2$; and the final value is obtained by extrapolation to $r^2=0$.
This procedure is reliable for the lowest six moments, $m=0,1,\ldots,5$ \cite{Li:2016dzv}.  The $m=5$ moment is not independent; but its direct calculation enables one to ensure that the lower even moments are correct.

One can extend this set of moments by using the Schlessinger point method (SPM) \cite{Schlessinger:1966zz, PhysRev.167.1411, Tripolt:2016cya, Chen:2018nsg, Binosi:2019ecz} to construct an analytic function, $M_S(z)$, whose values at $z=0,1,\ldots,5$ agree with the moments computed directly and for which $M_S(7)$ satisfies the constraint imposed by Eq.\,\eqref{xoneminusx}. The function $M_S(z)$ then provides an estimate for all moments of the distribution, which is exact for $m\leq 5$.

We tested the efficacy of this SPM approach using the algebraic model described in Ref.\,\cite{Xu:2018eii} (Eqs.\,(1), (14), (17) and Sec.IV.A).  Computing fifty Mellin moments directly, we then used the first six moments and the procedure described above to obtain a SPM approximation.  Comparing the moments obtained using the SPM approximation with the true moments, one finds the magnitude of the relative error is $<0.2\,$\% for $m\leq 10$ and $<1$\% for $m\leq 15$, \emph{i.e}.\ the SPM produces accurate approximations to the first sixteen moments, beginning with just six.

Having validated the SPM, we computed the moments in Eq.\,\eqref{MellinMoments} for $m=0,1,\ldots, 5$ using our numerical results for $S$ and $\Gamma_\pi$.  Then, to compensate for potential propagation of numerical quadrature error in the moment computations, we constructed two SPM approximations to the results: one based on the $m=0,1,2,3$ four-element subset; and another using the complete set of six moments.
Working with the first eleven SPM-approximant moments in each case, we reconstructed a pion valence-quark distribution; and subsequently defined our result to be the average of these functions:
\begin{align}
\nonumber {\mathpzc q}^\pi&(x;\zeta_H)  = 213.32 \, x^2 (1-x)^2\\
& \quad \times  [1 - 2.9342  \sqrt{x(1-x)} + 2.2911 \,x (1-x)]\,. \label{qpizetaH}
\end{align}
The mean absolute relative error between the first eleven moments computed using Eq.\,\eqref{qpizetaH} and those of the separate reconstructed distributions is 4(3)\%.
%

\begin{figure}[t]
\vspace*{0.5em}\leftline{\hspace*{2em}\textbf{A}}\vspace*{-2em}
\centerline{%
\includegraphics[clip, width=0.42\textwidth]{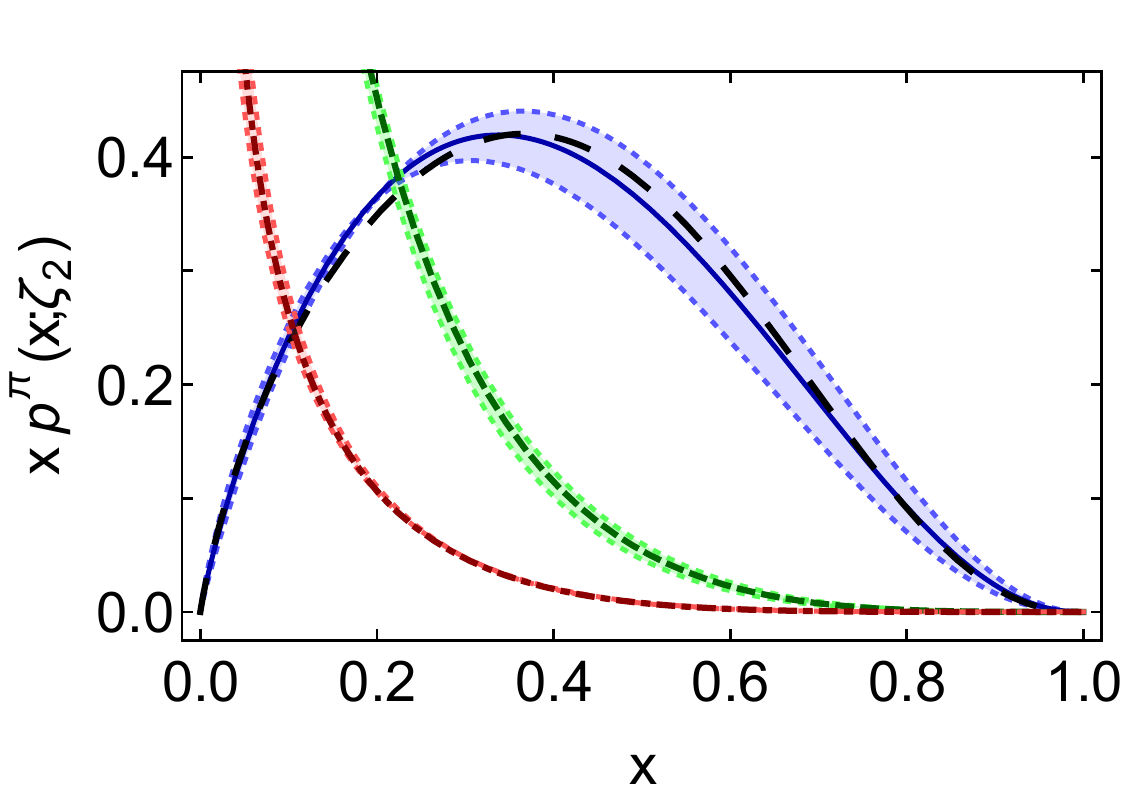}}
\vspace*{0.5em}\leftline{\hspace*{2em}\textbf{B}}\vspace*{-2em}
\centerline{%
\includegraphics[clip, width=0.42\textwidth]{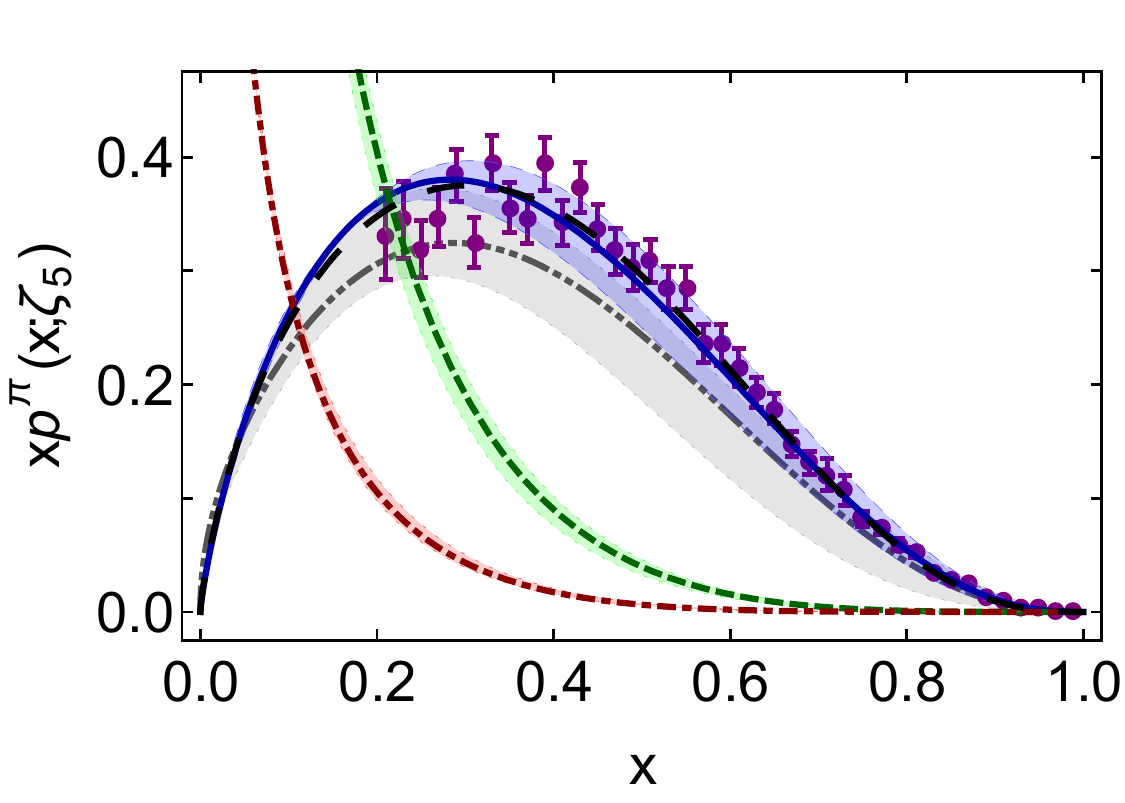}}
\caption{\label{qpizeta2}
Pion momentum distribution functions, $x p^\pi(x;\zeta)$, $p={\mathpzc q}, g, S$:
\textbf{A} (upper panel), evolved $\zeta_H \to \zeta_2=2\,$GeV; and
\textbf{B} (lower panel), evolved $\zeta_H \to \zeta_5=5.2\,$GeV.
Legend:
solid (blue) curve embedded in shaded band, our prediction for $x {\mathpzc q}(x;\zeta)$;
long-dashed (black) curve -- $\zeta_2$ result from Ref.\,\cite{Hecht:2000xa};
dashed (green) curve within shaded band -- predicted gluon distribution in pion, $x g^\pi(x;\zeta)$;
dot-dashed (red) curve within shaded band -- predicted sea-quark distribution, $x S^\pi(x;\zeta)$. (These last two are detailed in Eqs.\,\eqref{seaglueFunction}, \eqref{seagluefit}.)
In all the above cases, the shaded band indicates the effect of $\zeta_H \to \zeta_H (1 \pm 0.1)$.
\emph{Lower panel}: dot-dot-dashed (grey) curve within shaded band -- lQCD result \cite{Sufian:2019bol};
and data (purple) from Ref.\,\cite{Conway:1989fs}, rescaled according to the analysis in Ref.\,\cite{Aicher:2010cb}.
}
\end{figure}

Given the remarks in Sec.\,\emph{1}, it is worth highlighting that Eq.\,\eqref{qpizetaH} exhibits the $x\simeq 1$ behaviour predicted by the QCD parton model, Eq.\,\eqref{PDFQCD}; and because it is a purely valence distribution, this same behaviour is also evident on $x\simeq 0$.
However, in contrast to the scale-free valence-quark distribution computed in Ref.\,\cite{Chang:2014lva}:
$q_{\rm sf}(x) \approx 30 \, x^2 (1-x)^2$,
obtained using parton-model-like algebraic representations of $S$, $\Gamma_\pi$, the distribution computed with realistic inputs is a much broader concave function.  A similar effect is observed in the pion's leading-twist valence-quark distribution amplitude \cite{Chang:2013pq} and those of other mesons \cite{Gao:2014bca, Shi:2015esa, Gao:2016jka, Li:2016dzv, Li:2016mah}.  The cause is the same, \emph{viz}.\ the valence-quark distribution function is hardened owing to DCSB, which is a realisation of the mechanism responsible for the emergence of mass in the Standard Model \cite{Roberts:2016vyn}.  Emergent mass is expressed in the momentum-dependence of all QCD Schwinger functions.  It is therefore manifest in the pointwise behaviour of wave functions, elastic and transition form factors, \emph{etc}.; and as we have now displayed, also in parton distributions.

\smallskip

\noindent\emph{3.$\;$Evolution of pion distribution functions} ---
%
The pion valence-quark distribution in Eq.\,\eqref{qpizetaH} is computed at $\zeta_H=m_\alpha$.  On the other hand, existing lQCD calculations of low-order moments \cite{Best:1997qp, Detmold:2003tm, Brommel:2006zz, Oehm:2018jvm} and phenomenological fits to pion parton distributions are typically quoted at $\zeta \approx \zeta_2=2\,$GeV \cite{Sutton:1991ay, Gluck:1999xe, Barry:2018ort}; and the scale relevant to the E615 data is $\zeta_5=5.2\,$GeV \cite{Conway:1989fs, Wijesooriya:2005ir}.  We therefore employ leading-order QCD evolution of ${\mathpzc q}^\pi(x;\zeta_H=m_\alpha)$ to obtain results for ${\mathpzc q}^\pi(x;\zeta_2)$ and ${\mathpzc q}^\pi(x,\zeta_5)$ using the process-independent running coupling in Eq.\,\eqref{alphaPI}.   Notably, given that $\zeta_H=m_\alpha$ is fixed by our analysis, all results are predictions; and $\alpha_{\rm PI}(\zeta_H)/(2\pi)=0.20$, $[\alpha_{\rm PI}(\zeta_H)/(2\pi)]^2=0.04$, so that leading-order evolution serves as a good approximation.  We checked that with fixed $\zeta_H$, varying $m_\alpha \to (1 \pm 0.1) m_\alpha$ does not measurably affect the evolved distributions.  We therefore report results with $m_\alpha$ fixed and an uncertainty determined by varying $\zeta_H \to (1\pm0.1) \zeta_H$.

\begin{table}[t]
\caption{\label{fittingparameters}
Coefficients and powers that reproduce the computed pion valence-quark distribution functions, depicted in Fig.~\ref{qpizeta2}, when used in Eq.\,\eqref{PDFform}.
}
\begin{center}
\begin{tabular*}
{\hsize}
{
l@{\extracolsep{0ptplus1fil}}|
c@{\extracolsep{0ptplus1fil}}
c@{\extracolsep{0ptplus1fil}}
c@{\extracolsep{0ptplus1fil}}
c@{\extracolsep{0ptplus1fil}}
c@{\extracolsep{0ptplus1fil}}}\hline\hline
 & ${\mathpzc n}_{{\mathpzc q}^\pi}\ $ & $\alpha\ $ & $\beta\ $ & $\rho\ $ & $\gamma\ $ \\\hline
             & $9.83\ $ & $-0.080\ $ & $2.29\ $ & $-1.27\ $ & $0.511\ $  \\
$\zeta_2\ $ & $8.31\ $ & $-0.127\ $ & $2.37\ $ & $-1.19\ $ & $0.469\ $ \\
             & $7.01\ $ & $-0.162\ $ & $2.47\ $ & $-1.12\ $ & $0.453\ $ \\\hline
             & $7.81\ $ & $-0.153\ $ & $2.54\ $ & $-1.20\ $ & $0.505\ $ \\
$\zeta_5\ $ & $7.28\ $ & $-0.169\ $ & $2.66\ $ & $-1.21\ $ & $0.531\ $ \\
             & $6.48\ $ & $-0.188\ $ & $2.78\ $ & $-1.19\ $ & $0.555\ $  \\
\hline\hline
\end{tabular*}
\end{center}
\end{table}
%
%

Our prediction for ${\mathpzc q}^\pi(x;\zeta_2)$ is depicted in Fig.\,\ref{qpizeta2}A.  The solid (blue) curve and surrounding bands are described by the following function, a generalisation of Eq.\,\eqref{qpizetaH}:
\begin{align}
\nonumber {\mathpzc q}^\pi(x) & = {\mathpzc n}_{{\mathpzc q}^\pi} \,x^\alpha (1-x)^\beta \\
& \times [1 + \rho\, x^{\alpha/4} (1-x)^{\beta/4} + \gamma \,x^{\alpha/2} (1-x)^{\beta/2} ]\,,
\label{PDFform}
\end{align}
where ${\mathpzc n}_{{\mathpzc q}^\pi}$ ensures baryon number conservation and the powers and coefficients are listed in Table~\ref{fittingparameters}.  Evidently, the large-$x$ exponent is
$\beta(\zeta_2) = 2.38(9)$\,.

Here it is also worth listing an array of associated, calculated low-order moments in comparison with those obtained in the more recent lQCD simulations:
\begin{equation}
\label{momentslQCD}
\begin{array}{l|lll}
\zeta_2  & \langle x \rangle_u^\pi & \langle x^2 \rangle_u^\pi & \langle x^3 \rangle_u^\pi\\\hline
\mbox{Ref.\,\cite{Detmold:2003tm}} & 0.24(2) & 0.09(3) & 0.053(15)\\
\mbox{Ref.\,\cite{Brommel:2006zz}} & 0.27(1) & 0.13(1) & 0.074(10)\\
\mbox{Ref.\,\cite{Oehm:2018jvm}} & 0.21(1) & 0.16(3) & \\
\hline
{\rm Herein} & 0.24(2) & 0.098(10) & 0.049(07)
\end{array}\,.
\end{equation}
%
Both continuum and lQCD results agree on the light-front momentum fraction carried by valence-quarks in the pion at $\zeta=\zeta_2$:
$\langle 2 x \rangle_{\mathpzc q}^\pi =0.48(3)$,
\emph{i.e}.\ roughly one-half.  This is consistent with a recent phenomenological analysis of data on $\pi$-nucleus Drell-Yan and leading neutron electroproduction \cite{Barry:2018ort}:  $\langle 2 x \rangle_{\mathpzc q}^\pi  = 0.48(1)$ at $\zeta=2.24\,$GeV.

As explained above, the pion is purely a bound-state of a dressed-quark and dressed-antiquark at the hadronic scale: sea and glue distributions are zero at $\zeta_H$, being generated by QCD evolution on $\zeta>\zeta_H$.  Using LO evolution with the coupling in Eq.\,\eqref{alphaPI} we obtain the sea and glue distributions in Fig.\,\ref{qpizeta2}, from which one computes the following momentum fractions ($\zeta=\zeta_2$):
$\langle x\rangle^\pi_g = 0.41(2)$, 
$\langle x\rangle^\pi_{\rm sea} = 0.11(2)$.
The ordering of these values agrees with that in \cite{Barry:2018ort}, but our gluon momentum-fraction is $\sim 20$\% larger and that of the sea is commensurately smaller.

Our computed glue and sea momentum distributions are fairly approximated using the functional form:
\begin{equation}
\label{seaglueFunction}
x p^\pi(x;\zeta) = {\mathpzc A} \, x^\alpha \, (1-x)^\beta\,,
\end{equation}
with the coefficient and powers listed here ($p=g=\,$glue, $p=S=\,$sea):
\begin{equation}
\label{seagluefit}
\begin{array}{l|cccc}
             & p & {\mathpzc A} & \alpha & \beta \\\hline
\zeta_2 &  g & 0.40 \mp 0.03 & -0.55 \mp 0.03 & 3.47 \pm 0.13 \\
          &  S & 0.13 \mp 0.01 & -0.53 \mp 0.05 & 4.51 \pm 0.03 \\\hline
\zeta_5 &  g & 0.34 \mp 0.04 & -0.62 \mp 0.04 & 3.75 \pm 0.12 \\
          &  S & 0.12 \pm 0.02 & -0.61 \mp 0.07 & 4.77 \pm 0.03 \\\hline
\end{array}\,.
\end{equation}


Our predictions for the pion parton distributions at a scale relevant to the E615 experiment, \emph{i.e}.\ $\zeta_5=5.2\,$GeV \cite{Conway:1989fs, Wijesooriya:2005ir}, are depicted in Fig.\,\ref{qpizeta2}B.  The solid (blue) curve and surrounding bands are described by the function in Eq.\,\eqref{PDFform} with the powers and coefficients listed in Table~\ref{fittingparameters}.  Evidently, the large-$x$ exponent is
$\beta(\zeta_5) = 2.66(12)$\,.
Working with results obtained in an exploratory lQCD calculation \cite{Sufian:2019bol}, one finds $\beta_{\rm lQCD}(\zeta_5) = 2.45(58)$; and also the following comparison between low-order moments:
\begin{equation}
\label{momentslQCD5}
\begin{array}{l|lll}
\zeta_5  & \langle x \rangle_u^\pi & \langle x^2 \rangle_u^\pi & \langle x^3 \rangle_u^\pi\\\hline
\mbox{Ref.\,\cite{Sufian:2019bol}} & 0.17(1) & 0.060(9) & 0.028(7)\\
{\rm Herein} & 0.21(2) & 0.076(9) & 0.036(5)
\end{array} \,.
\end{equation}

The data in Fig.\,\ref{qpizeta2}B is that reported in Ref.\,\cite{Conway:1989fs}, rescaled according to the analysis in Ref.\,\cite{Aicher:2010cb}.  Our prediction agrees with the rescaled data.  Importantly, no parameters were varied in order to achieve this outcome, or any other reported herein.

As above, the predictions for the glue and sea distributions in Fig.\,\ref{qpizeta2}B were obtained using  LO evolution from $\zeta_H=m_\alpha \to \zeta_5$ with the coupling in Eq.\,\eqref{alphaPI}; and from these distributions one obtains the following momentum fractions ($\zeta=\zeta_5$):
$\langle x\rangle^\pi_g = 0.45(1)$, 
$\langle x\rangle^\pi_{\rm sea} = 0.14(2)$.
The glue and sea momentum distributions are fairly described by the function in Eq.\,\eqref{seaglueFunction} evaluated using the coefficient and powers in the lower rows of Eq.\,\eqref{seagluefit}.  (Recall that on $\Lambda_{\rm QCD}^2/\zeta^2 \simeq 0$, for any hadron \cite{Altarelli:1981ax}: $\langle x\rangle_q =0$, $\langle x\rangle_g = 4/7\approx 0.57$, $\langle x\rangle_S = 3/7\approx 0.43$.)

Figure~\ref{qpizeta2}B also displays the lQCD result for the pion valence-quark distribution function \cite{Sufian:2019bol} evolved to the E615 scale: dot-dot-dashed (grey) curve within bands.  As could be  anticipated from the comparisons listed in connection with Eq.\,\eqref{momentslQCD5}, the pointwise form of the lQCD prediction agrees with our result (within errors).  

\smallskip

\noindent\emph{4.$\;$Perspective} ---
%
%
Our symmetry-preserving analysis of the pion's parton distribution functions exploits the existence of a process-independent effective charge in QCD, which saturates at infrared momenta \cite{Binosi:2016nme, Rodriguez-Quintero:2018wma}, to introduce an unambiguous definition of the hadronic scale, $\zeta_H$, and thereby obtain parameter-free predictions, unified with kindred results for the electromagnetic pion elastic and transition form factors
\cite{Chang:2013nia, Raya:2015gva, Raya:2016yuj, Gao:2017mmp, Chen:2018rwz, Ding:2018xwy} and numerous other observables (\emph{e.g}.\ Refs.\,\cite{Binosi:2018rht, Qin:2019hgk}).
At $\zeta_H$, the computed valence-quark distribution is \emph{hard}, as a direct consequence of DCSB, \emph{i.e}.\ the mechanism which expresses the emergence of mass in the fermion sector of QCD.

Evolved to $\zeta=5.2\,$GeV, the calculated distribution agrees with that obtained in a recent, exploratory lattice-QCD computation \cite{Sufian:2019bol}.  With this confluence, two disparate treatments of the pion bound-state problem are seen to have arrived at the \emph{same} prediction for the pion's valence-quark distribution function.  It also agrees with $\pi$-nucleon Drell-Yan data \cite{Conway:1989fs}, rescaled as suggested by the complete next-to-leading-order (NLO) reanalysis in Ref.\,\cite{Aicher:2010cb}.  Importantly, via evolution, we also deliver realistic predictions for the pion's glue and sea content.

A pressing extension of this study is the calculation of analogous kaon distribution functions, which would enable validation of earlier analyses \cite{Chen:2016sno} that indicate the kaon's gluon content is significantly smaller than that of the pion owing to DCSB and its role in forming the almost-massless pion \cite{Roberts:2016vyn}.

\begin{acknowledgments}
%
%
We are grateful for constructive comments from C.~Chen, M.~Chen, F.~Gao, C.~Mezrag, J.~Papavassiliou, J.~Repond, J.~Rodr{\'{\i}}guez-Quintero and J.~Segovia;
for the hospitality and support of RWTH Aachen University, III.\,Physikalisches Institut B, Aachen - Germany;
and likewise for the hospitality and support of the University of Huelva, Huelva - Spain, and the University of Pablo de Olavide, Seville - Spain, during the ``4th Workshop on Nonperturbative QCD'' (University of Pablo de Olavide, 6-9 November 2018).
Work supported by:
the Chinese Government's \emph{Thousand Talents Plan for Young Professionals};
Jiangsu Province \emph{Hundred Talents Plan for Professionals};
and Forschungszentrum J\"ulich GmbH.
\end{acknowledgments}



\providecommand{\newblock}{}

\end{document}